\documentclass[11pt]{article}
\usepackage{amssymb,amsmath,amsfonts}
\usepackage{graphicx}
\usepackage{graphics}
\usepackage{eepic,epsfig}

\begin{document}

\title{ Coherent bremsstrahlung in periodically deformed crystals \\ with a complex base}
\author{A. R. Mkrtchyan, A. A. Saharian\thanks{%
E-mail: saharian@ictp.it}, V. V. Parazian \\
\textit{\small Institute of Applied Problems in Physics, 25
Nersessian Street, 0014 Yerevan, Armenia}} \maketitle

\begin{abstract}
In the present paper we investigate coherent bremsstrahlung of
high energy electrons moving in a periodically deformed single
crystal with a complex base. The formula for corresponding
differential cross-section is derived for an arbitrary deformation
field. The conditions are discussed under which the influence of
the deformation is important. The case is considered in detail
when the electron enters into the crystal at small angles with
respect to a crystallographic axis. It is shown that in dependence
of the parameters, the presence of the deformation can either
enhance or reduce the bremsstrahlung cross-section.
\end{abstract}

\bigskip


\bigskip


\bigskip

\section{Introduction}

\label{sec1}

The high-energy electromagnetic processes in crystals have been under active
theoretical investigations for long time (see, for instance, \cite{TerMik}-%
\cite{Uger05} and references therein). These investigations are of interest
not only from the viewpoint of underlying physics but also from the
viewpoint of practical applications. When the formation length exceeds the
interatomic spacing, the interference effects from all atoms within this
length are important and the cross-sections of the electromagnetic processes
in crystals can change essentially compared with the corresponding
quantities for an isolated atom. From the point of view of controlling the
parameters of the high-energy electromagnetic processes in a medium it is of
interest to investigate the influence of external fields, such as acoustic
waves, temperature gradient, etc., on the corresponding characteristics. The
considerations of concrete processes, such as the diffraction radiation,
transition radiation, parametric X-radiation, channelling radiation,
electron-positron pair creation by high-energy photons, have shown that the
external fields can essentially change the angular-frequency characteristics
of these processes (see, for example \cite{MkrtDR}-\cite{Mkrt05}).

The coherent bremsstrahlung of high-energy electrons moving in a
crystal is one of the most effective methods to produce intense
beams of highly polarized and monochromatic photons (for recent
experiments in this direction see, for instance, \cite{Apya08} and
references therein). Such radiation has a number of remarkable
properties and at present it has found many important
applications. This motivates the importance of investigations for
various mechanisms of controlling the radiation parameters. As
such a mechanism, in \cite{Mkrtbrem}\ we have discussed the
influence of hypersonic waves excited in a crystal on the process
of the bremsstrahlung of high-energy electrons. In this paper, the
case of a simplest crystal with a single atom in the elementary
base and the sinusoidal deformation field generated by the
hypersound were considered. To have an essential influence of the
acoustic wave, high-frequency hypersound is needed. Usually this
type of waves are excited by high-frequency electromagnetic fields
through the piezoelectric effect in crystals with a complex
elementary base. In the present paper we generalize the results of
\cite{Mkrtbrem,Para06} for crystals with a complex base and for
acoustic waves with an arbitrary profile. The numerical
calculations are carried out for the case of the most popular
quartz piezocrystal.

The paper is organized as follows. In the next section we derive the general
formula for the coherent part of the bremsstrahlung cross-section averaged
over the thermal fluctuations and the conditions are specified under which
the influence of the deformation field can be considerable. The analysis of
the general formula in the case when the electron enters into the crystal at
small angles with respect to the crystallographic axes or planes is given in
section \ref{sec3}. The results of the numerical calculations for the
cross-section as a function of the photon energy and the amplitude of the
external excitation are presented. Section \ref{sec4} summarizes the main
results of the paper. Throughout of the paper the system of units $\hbar
=c=1\ $is used.

\section{Influence of external excitations on bremsstrahlung in crystals}

\label{sec2}

We consider the bremsstrahlung of high energy electrons moving in a crystal.
Let $\left( \omega ,\mathbf{k}\right) $ be the frequency and the wave vector
for the radiated photon and $\left( E_{1,}\mathbf{p}_{1}\right) $, $\left(
E_{2,}\mathbf{p}_{2}\right) $ be the energies and momenta for the electrons
in the initial and final states respectively. We denote by $d^{4}\sigma
_{0}/d\omega d^{3}q=\big|u_{\mathbf{q}}^{\left( j\right) }\big|^{2}\sigma
_{0}\left( \mathbf{q}\right) $ the cross-section of the bremsstrahlung on an
isolated $j$-th atom as a function of the transfer momenta $\mathbf{q=p}_{1}-%
\mathbf{p}_{2}-\mathbf{k}$. In this representation $u_{\mathbf{q}}^{\left(
j\right) }$ is the Fourier transform of the potential for $j$-th atom and
the factor $\sigma _{0}\left( \mathbf{q}\right) $ does not depend on the
type of the atom. The quantity $u_{\mathbf{q}}^{\left( j\right) }$ usually
is presented in the form $4\pi Z_{j}e^{2}\left[ 1-F^{\left( j\right) }\left(
q\right) \right] /q^{2}$, where $Z_{j}$ and $F^{\left( j\right) }\left(
q\right) $ are the number of electrons in the atom and the atomic
form-factor. The differential cross-section for the bremsstrahlung in a
crystal by high energy electron can be written in the form (see, for example
\cite{TerMik,Shulga})%
\begin{equation}
\sigma \left( \mathbf{q}\right) \equiv \frac{d^{4}\sigma }{d\omega d^{3}q}=%
\bigg|\sum_{n,j}u_{\mathbf{q}}^{\left( j\right) }e^{i\mathbf{qr}_{n}}\bigg|%
^{2}\sigma _{0}\left( \mathbf{q}\right) ,  \label{sigmadefin}
\end{equation}%
where the vector $\mathbf{r}_{n}^{\left( j\right) }$ specifies the positions
of the atoms in the crystal and the collective index $n$\ numerates
elementary cells.

At nonzero temperature one can decompose the radius-vector as $\mathbf{r}%
_{n}^{\left( j\right) }=\mathbf{r}_{n0}^{\left( j\right) }+\mathbf{u}%
_{tn}^{\left( j\right) }$, where $\mathbf{u}_{tn}^{\left( j\right) }$ is the
displacement of $j$-th atom with respect to the equilibrium position $%
\mathbf{r}_{n0}^{\left( j\right) }$\ due to the thermal vibrations. After
averaging over thermal fluctuations, we can write cross-section as (see for
example, \cite{TerMik,Shulga} for the case of a crystal with simple cell)%
\begin{equation}
\sigma \left( \mathbf{q}\right) =\bigg[N\sum_{j}\big|u_{\mathbf{q}}^{\left(
j\right) }\big|^{2}\Big(1-e^{-q^{2}\overline{u_{t}^{\left( j\right) 2}}}\Big)%
+|U_{\mathbf{q}}|^{2}\bigg]\sigma _{0}\left( \mathbf{q}\right) ,
\label{sigTerMik}
\end{equation}%
where $N$ is the number of cells, $\overline{u_{t}^{\left( j\right) 2}}$ is
the temperature dependent mean-squared amplitude of the thermal vibrations
for the $j$-th atom, and we have introduced the notation%
\begin{equation}
U_{\mathbf{q}}=\sum_{n,j}u_{\mathbf{q}}^{\left( j\right) }e^{i\mathbf{qr}%
_{n0}^{\left( j\right) }}e^{-\frac{1}{2}q^{2}\overline{u_{t}^{\left(
j\right) 2}}},  \label{Uq}
\end{equation}%
with $e^{-q^{2}\overline{u_{t}^{\left( j\right) 2}}/2}$ being the
corresponding Debye-Waller factor. In formula (\ref{sigTerMik}) the first
term in the square brackets does not depend on the direction of the momentum
transfer $\mathbf{q}$ and it determines the contribution of the incoherent
effects. The contribution of the coherent effects is presented by the second
term which depends on the orientation of the crystal axes with respect to
the vector $\mathbf{q}$. By taking into account the formula for the
cross-section on an isolated atom, in the range of the transferred momenta $%
q\ll m_{e}$, with $m_{e}$ being the mass of electron, the coherent part of
the cross-section can be presented in the form%
\begin{equation}
\sigma _{c}\left( \mathbf{q}\right) =\frac{e^{2}}{8\pi ^{3}E_{1}^{2}}\frac{%
q_{\bot }^{2}}{q_{\Vert }^{2}}\bigg(1+\frac{\omega \delta }{m_{e}^{2}}-\frac{%
2\delta }{q_{\Vert }}+\frac{2\delta ^{2}}{q_{\Vert }^{2}}\bigg)|U_{\mathbf{q}%
}|^{2},  \label{sigcohdef}
\end{equation}%
where $q_{\Vert }\geqslant \delta $. In formula (\ref{sigcohdef}), $e$ is
the electron charge, $q_{\Vert }$ and $\mathbf{q}_{\bot }$ are the parallel
and perpendicular components of the vector $\mathbf{q}$ with respect to the
direction of the electron initial momentum $\mathbf{p}_{1}$, $\delta
=1/l_{c} $ is the minimum longitudinal momentum transfer, and $%
l_{c}=2E_{1}E_{2}/\left( \omega m_{e}^{2}\right) $ is the formation length
for the process of bremsstrahlung. The latter determines the effective
longitudinal dimension of the interaction region for the phase coherence of
the radiation process

When external influences are present (for example, in the form of acoustic
waves) the radius-vector of an atom in the crystal can be written as $%
\mathbf{r}_{n0}^{\left( j\right) }=\mathbf{r}_{ne}^{\left( j\right) }+%
\mathbf{u}_{n}^{\left( j\right) }$, where $\mathbf{r}_{ne}^{\left( j\right)
} $ determines the equilibrium positions of the atom in the situation
without the deformation, $\mathbf{u}_{n}^{\left( j\right) }$ is the
displacement of the atom caused by the external influence. In this paper we
will consider deformations having the periodical structure:%
\begin{equation}
\mathbf{u}_{n}^{\left( j\right) }=\mathbf{u}_{0}f(\mathbf{k}_{s}\mathbf{r}%
_{ne}^{\left( j\right) }),  \label{displace}
\end{equation}%
where $\mathbf{u}_{0}$ and $\mathbf{k}_{s}$ are the amplitude and the wave
vector of the deformations field, $f\left( x\right) $ is an arbitrary
function with the period $2\pi $, $\max f\left( x\right) =1$. In the
discussion below we will assume that $f\left( x\right) \in C^{\infty }\left(
R\right) $. The time-dependence of $\mathbf{u}_{n}^{\left( j\right) }$ in
the case of acoustic waves we can disregard, as for the electron energies we
are interested in, the characteristic time for the change of the deformation
field is much greater than the time of passage of the particles through the
crystal. For the deformation field given by Eq. (\ref{displace}),
introducing the Fourier transform of the function $e^{ixf\left( t\right) }$,
\begin{equation}
F_{m}\left( x\right) =\frac{1}{2\pi }\int_{-\pi }^{+\pi }e^{ixf\left(
t\right) -imt}dt,  \label{Fmxdef}
\end{equation}%
the sum over $n$ in (\ref{Uq}) is presented in the form%
\begin{equation}
\sum_{n}u_{\mathbf{q}}^{\left( j\right) }e^{i\mathbf{qr}_{n0}^{\left(
j\right) }}=\sum_{m=-\infty }^{+\infty }F_{m}\left( \mathbf{qu}_{0}\right)
\sum_{n}u_{\mathbf{q}}^{\left( j\right) }e^{i\mathbf{q}_{m}\mathbf{r}%
_{ne}^{\left( j\right) }},  \label{displaceFm}
\end{equation}%
where $\mathbf{q}_{m}=\mathbf{q+}m\mathbf{k}_{s}$.

For a lattice with a complex cell the coordinates of the atoms can be
presented in the form $\mathbf{r}_{ne}^{\left( j\right) }=\mathbf{R}_{n}+%
\mathbf{\rho }^{\left( j\right) }$, where $\mathbf{R}_{n}$ determines the
positions of the atoms of one of the primitive lattices, and $\mathbf{\rho }%
^{\left( j\right) }$ are the equilibrium positions for the other atoms
inside $n$-th elementary cell with respect to $\mathbf{R}_{n}$. Now the
Fourier transform of the effective potential can be written as%
\begin{equation}
U_{\mathbf{q}}=\sum_{m=-\infty }^{+\infty }F_{m}\left( \mathbf{qu}%
_{0}\right) S\left( \mathbf{q,q}_{m}\right) \sum_{n}e^{i\mathbf{q}_{m}%
\mathbf{R}_{n}},  \label{Uq1}
\end{equation}%
where the factor $S\left( \mathbf{q,q}_{m}\right) $ determined by structure
of the elementary cell is given by the formula%
\begin{equation}
S\left( \mathbf{q,q}_{m}\right) =\sum_{j}u_{\mathbf{q}}^{\left( j\right)
}e^{-\frac{1}{2}q^{2}\overline{u_{t}^{\left( j\right) 2}}}e^{i\mathbf{q}_{m}%
\mathbf{\rho }^{\left( j\right) }}.  \label{strctfacdef}
\end{equation}

In the case of thick crystals, the sum over the cells in (\ref{Uq1}) is
expressed in terms of a sum over the reciprocal lattice and one finds%
\begin{equation}
U_{\mathbf{q}}=\frac{\left( 2\pi \right) ^{3}}{\Delta }\sum_{m=-\infty
}^{+\infty }F_{m}\left( \mathbf{qu}_{0}\right) S\left( \mathbf{q,q}%
_{m}\right) \sum_{\mathbf{g}}\delta \left( \mathbf{q}-\mathbf{g}_{m}\right) ,
\label{Uq2}
\end{equation}%
where $\Delta $ is the volume of the unit cell, $\mathbf{g}$ is the
reciprocal lattice vector, and we have introduced the notation%
\begin{equation}
\mathbf{g}_{m}=\mathbf{g-}m\mathbf{k}_{s}.  \label{gm}
\end{equation}%
Note that the function $U_{\mathbf{q}}$, given by formula (\ref{Uq2}), is
the Fourier transform of the periodic function
\begin{equation}
U(\mathbf{r})=\,\frac{1}{\Delta }\sum_{m,\mathbf{g}}F_{m}\left( \mathbf{g}%
_{m}\mathbf{u}_{0}\right) S\left( \mathbf{g}_{m}\mathbf{,g}\right) e^{-i%
\mathbf{g}_{m}\mathbf{r}}.  \label{EffPot}
\end{equation}%
Thus, in the presence of the periodic deformation field (\ref{displace}),
the process of coherent bremsstrahlung can be considered as a result of the
electron motion in the continuous periodic potential given by (\ref{EffPot}%
). This field introduces additional momentum to the problem changing the
kinematics. By taking into account the delta-function in formula (\ref{Uq2}%
), the momentum conservation law is written in the form%
\begin{equation}
\mathbf{p}_{1}=\mathbf{p}_{2}+\mathbf{k+g-}m\mathbf{k}_{s},
\label{momentconserv}
\end{equation}%
where $-m\mathbf{k}_{s}$ is the momentum transfer to the external field.

For thick crystals, by using (\ref{Uq2}), for the factor in the coherent
part of the cross-section we have%
\begin{eqnarray}
|U_{\mathbf{q}}|^{2} &=&\frac{\left( 2\pi \right) ^{3}}{\Delta }\sum_{m,%
\mathbf{g}}F_{m}\left( \mathbf{g}_{m}\mathbf{u}_{0}\right) S\left( \mathbf{g}%
_{m}\mathbf{,g}\right) \delta \left( \mathbf{q}-\mathbf{g}_{m}\right)  \notag
\\
&&\times \sum_{m^{\prime }}F_{m^{\prime }}^{\ast }\left( \mathbf{qu}%
_{0}\right) S^{\ast }\left( \mathbf{q,q}_{m^{\prime }}\right)
\sum_{n}e^{-i\left( m-m\prime \right) \mathbf{k}_{s}\mathbf{R}_{n}}.
\label{UqMod2}
\end{eqnarray}%
Under the assumptions for the function $f\left( x\right) $ given above, by
making use the stationary phase method for the integral (\ref{Fmxdef}), we
can see that for a fixed $x$ one has $F_{m}\left( x\right) \sim \mathcal{O}%
\left( \left\vert m\right\vert ^{-\infty }\right) $ for $m\rightarrow \infty
$. Using this property, in the way similar to that used in \cite{MkrtParpair}%
, we can see that in the sum over $m$ the main contribution comes from the
terms for which $\left\vert m\mathbf{k}_{s}\mathbf{u}_{0}\right\vert
\lesssim \left\vert \mathbf{gu}_{0}\right\vert $, or $\left\vert
m\right\vert \lesssim \lambda _{s}/a$, where $\lambda _{s}=2\pi /k_{s}$ is
the wavelength of the external excitation, and $a$\ is of the order of the
lattice constants. Under the condition \ $u_{0}/\lambda _{s}\ll 1$, in the
summation on the right of formula (\ref{UqMod2}) the contributions of the
terms with $m\neq m^{\prime }$ are small compared to the diagonal terms $%
m=m^{\prime }$ and the sum over $n$ is equal to the number of the cells $N$
in the crystal. As a result we find%
\begin{equation}
|U_{\mathbf{q}}|^{2}=N\frac{\left( 2\pi \right) ^{3}}{\Delta }\sum_{m,%
\mathbf{g}}\left\vert F_{m}\left( \mathbf{g}_{m}\mathbf{u}_{0}\right)
\right\vert ^{2}\left\vert S\left( \mathbf{g}_{m},\mathbf{g}\right)
\right\vert ^{2}\delta \left( \mathbf{q}-\mathbf{g}_{m}\right) .
\label{UqMod21}
\end{equation}

Substituting the expression (\ref{UqMod21}) into the formula (\ref{sigcohdef}%
) and integrating over the vector $\mathbf{q}$, for the cross-section one
obtains%
\begin{equation}
d\sigma =\int \sigma \left( \mathbf{q}\right) d\mathbf{q}=N_{0}\left(
d\sigma _{n}+d\sigma _{c}\right) ,  \label{sigcohnoncoh}
\end{equation}%
with $d\sigma _{n}$ and $d\sigma _{c}$ being the incoherent and coherent
parts of the cross-section per atom, $N_{0}$ is the number of atoms in the
crystal. The coherent part of the cross-section is determined by the formula%
\begin{equation}
\frac{d\sigma _{c}}{d\omega }=\frac{e^{2}N}{N_{0}E_{1}^{2}\Delta }\sum_{m,%
\mathbf{g}}\frac{g_{m\bot }^{2}}{g_{m\Vert }^{2}}\left[ 1+\frac{\omega ^{2}}{%
2E_{1}E_{2}}-2\frac{\delta }{g_{m\parallel }}\left( 1-\frac{\delta }{%
g_{m\parallel }}\right) \right] \left\vert F_{m}\left( \mathbf{g}_{m}\mathbf{%
u}_{0}\right) \right\vert ^{2}\left\vert S\left( \mathbf{g}_{m},\mathbf{g}%
\right) \right\vert ^{2},  \label{sigcohgener}
\end{equation}%
where the vector $\mathbf{g}_{m}$ is defined by relation (\ref{gm}) and the
summation is carried out under the condition $g_{m\parallel }\geqslant
\delta $. For monoatomic single crystals and for sinusoidal deformation
fields, $f\left( z\right) =\sin \left( z+\varphi _{0}\right) $, the formula (%
\ref{sigcohgener}) is reduced to the result obtained in \cite{Mkrtbrem}.
Note that for this type of deformation one has the Fourier-transform%
\begin{equation}
F_{m}\left( z\right) =e^{im\varphi _{0}}J_{m}\left( z\right)
\label{FmBessel}
\end{equation}%
with $J_{m}\left( z\right) $\ being the Bessel function. The formula (\ref%
{sigcohgener}) differs from the corresponding expression for the
bremsstrahlung in undeformed crystals by the replacement $\mathbf{%
g\longrightarrow g}_{m}$, and by an additional summation over $m$\ with the
weights $\left\vert F_{m}\left( \mathbf{g}_{m}\mathbf{u}_{0}\right)
\right\vert ^{2}$. This corresponds to the presence of an additional
one-dimensional superlattice with the period $\lambda _{s}$ and with the
reciprocal lattice vector $m\mathbf{k}_{s}$, $m=0,\pm 1,\pm 2,...$. In the
presence of the deformation field the number of possibilities to satisfy the
condition $g_{m\parallel }\geqslant \delta $ in the summation of formula (%
\ref{sigcohgener}) increases due to the term $mk_{s\Vert }$ in the
expression for $g_{m\parallel }$. As we will see below, this leads to the
appearance of additional peaks in the spectral distribution of the radiated
photons. As the main contribution into the coherent part of the
cross-section comes from the terms with $g_{m\Vert }\sim \delta $, the
influence of the deformation field may be considerable if $\left\vert
mk_{s\Vert }\right\vert \gtrsim \delta $. Combining this with the previous
estimate that the main contribution into the series over $m$ comes from the
terms $\left\vert m\right\vert \lesssim \lambda _{s}/a$, we find the
following condition: $u_{0}/\lambda _{s}\gtrsim a/(4\pi ^{2}l_{c})$. At high
energies one has $a/l_{c}\ll 1$ and this condition can be consistent with
the condition $u_{0}/\lambda _{s}\ll 1$.

\section{Limiting cases and numerical results}

\label{sec3}

In this section we consider the general formula
(\ref{sigcohgener}) for the cross-section of the coherent part of
the bremsstrahlung in special cases when the coherence effects are
important. We will assume that the crystal lattice is
orthogonal. The reciprocal lattice vector components are given by $%
g_{i}=2\pi n_{i}/a_{i}$, $n_{i}=0,\pm 1,\pm 2,...$, where $a_{i}$, $i=1,2,3$%
, are the lattice constants along the corresponding directions. Let $\theta $
be the angle between the initial electron momentum $\mathbf{p}_{1}$ and the
crystallographic $z$-axis. For the parallel component of the vector $\mathbf{%
g}_{m}$ we have the expression%
\begin{equation}
g_{m\Vert }=g_{mz}\cos \theta +\left( g_{my}\cos \alpha +g_{mx}\sin \alpha
\right) \sin \theta ,  \label{gmgmzgmygmx}
\end{equation}%
where $\alpha $ is the angle between the projection of the vector $\mathbf{p}%
_{1}$ on the plane $\left( x,y\right) $ and $y$-axis. If the electron moves
in an unoriented crystal, in formula (\ref{sigcohgener}) the summation over $%
\mathbf{g}$ can be replaced by the integration and the cross-section for the
bremsstrahlung coincides with that in an amorphous medium. Coherent effects
appear when the electron enters into the crystal at small angles $\theta $.
In this case the main contribution to the cross-section give the terms with $%
g_{z}=0$, and from formula (\ref{sigcohgener}) one finds%
\begin{equation}
\frac{d\sigma _{c}}{d\omega }\approx \frac{e^{2}N}{E_{1}^{2}N_{0}\Delta }%
\sum_{m,g_{x},g_{y}}\frac{g_{m\bot }^{2}}{g_{m\Vert }^{2}}\left[ 1+\frac{%
\omega ^{2}}{2E_{1}E_{2}}-2\frac{\delta }{g_{m\parallel }}\left( 1-\frac{%
\delta }{g_{m\parallel }}\right) \right] \left\vert F_{m}\left( \mathbf{g}%
_{m}\mathbf{u}_{0}\right) \right\vert ^{2}\left\vert S\left( \mathbf{g}_{m},%
\mathbf{g}\right) \right\vert ^{2},  \label{sigcohsumgxgy}
\end{equation}%
where the summation goes under the condition $g_{m\parallel }\geqslant
\delta $ with%
\begin{equation}
g_{m\Vert }\approx -mk_{s\parallel }+\left( g_{y}\cos \alpha +g_{x}\sin
\alpha \right) \theta .  \label{gmmksgygx}
\end{equation}%
Note that in the arguments of the functions $F_{m}$ and $S$ we can put $%
g_{m}\approx \left( g_{x},g_{y},0\right) $.

In the further discussion two qualitatively different cases should be
distinguished. The first one corresponds to the situation where the electron
moves far from the crystallographic planes (angles $\alpha $ and $\pi
/2-\alpha $ are not small). In this case, the summation over the reciprocal
lattice vector components $g_{x}$ and $g_{y}$ in (\ref{sigcohsumgxgy}) can
be replaced by the integration in accordance with $\sum_{g_{x},g_{y}}%
\rightarrow \left( a_{1}a_{2}/4\pi ^{2}\right) \int dg_{x}g_{y}$. For the
corresponding cross-section this leads to the result
\begin{eqnarray}
\frac{d\sigma _{c}}{d\omega } &\approx &\frac{e^{2}N}{4\pi
^{2}E_{1}^{2}a_{3}N_{0}}\sum_{m}\int dg_{x}dg_{y}\frac{g_{m\bot }^{2}}{%
g_{m\Vert }^{2}}\left\vert F_{m}\left( \mathbf{g}_{m}\mathbf{u}_{0}\right)
\right\vert ^{2}  \notag \\
&&\times \left\vert S\left( \mathbf{g}_{m},\mathbf{g}\right) \right\vert ^{2}%
\left[ 1+\frac{\omega ^{2}}{2E_{1}E_{2}}-2\frac{\delta }{g_{m\parallel }}%
\left( 1-\frac{\delta }{g_{m\parallel }}\right) \right] ,
\label{sigcohintegrgxgy}
\end{eqnarray}%
where the integration range is determined by the condition $g_{m\parallel
}\geqslant \delta $ with $g_{m\Vert }$ given by (\ref{gmmksgygx}).

In the second case the electron enters into the crystal at small angles $%
\theta $ with respect to the crystallographic axis $z$ and near the
crystallographic planes $\left( y,z\right) $ (the angle $\alpha $ is small).
In this case, in dependence of the electron energy, two subcases should be
considered separately. Under the condition $\delta \sim 2\pi \theta /a_{2}$,
for the longitudinal component in formula (\ref{sigcohsumgxgy}) one has%
\begin{equation}
g_{m\Vert }\approx -mk_{s\parallel }+\theta g_{y}\geqslant \delta ,
\label{gmmkzgy}
\end{equation}%
and the summation over $g_{x}$ can be replaced by the
integration, $\sum_{g_{x}}\rightarrow \left( a_{1}/2\pi \right) \int dg_{x}$%
, with the result%
\begin{eqnarray}
\frac{d\sigma _{c}}{d\omega } &\approx &\frac{e^{2}N}{2\pi
E_{1}^{2}a_{2}a_{3}N_{0}}\sum_{m,g_{y}}\int dg_{x}\frac{g_{m\bot }^{2}}{%
g_{m\Vert }^{2}}\left\vert F_{m}\left( \mathbf{g}_{m}\mathbf{u}_{0}\right)
\right\vert ^{2}  \notag \\
&&\times \left\vert S\left( \mathbf{g}_{m},\mathbf{g}\right) \right\vert ^{2}%
\left[ 1+\frac{\omega ^{2}}{2E_{1}E_{2}}-2\frac{\delta }{g_{m\parallel }}%
\left( 1-\frac{\delta }{g_{m\parallel }}\right) \right] .
\label{sigcohintegrgxsumgy}
\end{eqnarray}%
This formula is further simplified in the case when the amplitude of the
deformation field, $\mathbf{u}_{0}$, is perpendicular to the
crystallographic $x$-axis. In this case, in the argument of the function $%
F_{m}$ in (\ref{sigcohintegrgxsumgy}) one has $\mathbf{g}_{m}\mathbf{u}%
_{0}\approx g_{y}u_{0y}$ and this function does not depend on the
integration variable. As a result, for the cross-section we obtain
the formula
\begin{eqnarray}
\frac{d\sigma _{c}}{d\omega } &\approx &\frac{e^{2}N}{2\pi
E_{1}^{2}a_{2}a_{3}N_{0}}\sum_{m,g_{y}}\left[ 1+\frac{\omega ^{2}}{%
2E_{1}E_{2}}-2\frac{\delta }{g_{m\parallel }}\left( 1-\frac{\delta }{%
g_{m\parallel }}\right) \right]  \notag \\
&&\times \frac{\left\vert F_{m}\left( \mathbf{g}_{m}\mathbf{u}_{0}\right)
\right\vert ^{2}}{g_{m\Vert }^{2}}\int dg_{x}g_{\bot }^{2}\left\vert S\left(
\mathbf{g}_{m},\mathbf{g}\right) \right\vert ^{2},
\label{sigsumgyintgxstruc}
\end{eqnarray}%
with an effective structure factor determined by the integral on the
right-hand side.

In the second subcase we assume the electron energies for which $\delta \sim
2\pi \theta \alpha /a_{1}$. Now, the main contribution into the sum in Eq. (%
\ref{sigcohsumgxgy}) comes from the terms with $g_{y}=0$, and the formula
for the cross-section will become%
\begin{equation}
\frac{d\sigma _{c}}{d\omega }\approx \frac{e^{2}N}{E_{1}^{2}N_{0}\Delta }%
\sum_{m,g_{x}}\frac{g_{m\bot }^{2}}{g_{m\Vert }^{2}}\left[ 1+\frac{\omega
^{2}}{2E_{1}E_{2}}-2\frac{\delta }{g_{m\parallel }}\left( 1-\frac{\delta }{%
g_{m\parallel }}\right) \right] \left\vert F_{m}\left( \mathbf{g}_{m}\mathbf{%
u}_{0}\right) \right\vert ^{2}\left\vert S\left( \mathbf{g}_{m},\mathbf{g}%
\right) \right\vert ^{2},  \label{sigsummn1}
\end{equation}%
where one has
\begin{equation}
g_{m\Vert }\approx -mk_{z\parallel }+\psi g_{x},\text{ \ }\psi \equiv \alpha
\theta ,  \label{gmmkpsigx}
\end{equation}%
and, as before, the summation goes under the condition $g_{m\parallel
}\geqslant \delta $.

We have performed numerical calculations for the bremsstrahlung
cross-section for various values of parameters in the case of $\mathrm{SiO}%
_{2}$ single crystal at zero temperature. The corresponding results show
that by adjusting the orientation of the crystal relative to the incident
electron momentum and the parameters of the external influence it is
possible to enhance the number of bremsstrahlung photons. To deal with an
orthogonal lattice, as an elementary cell we choose the cell including 6
atoms of silicon and 12 atoms of oxygen (elementary cell of Shrauf). For the
Fourier transforms of the atomic potentials we take the Moliere
parametrization:%
\begin{equation}
u_{q}^{\left( j\right) }=\sum_{i=1}^{3}\frac{4\pi Z_{j}e^{2}\alpha _{i}}{%
q^{2}+\left( \chi _{i}/R_{j}\right) ^{2}},  \label{molierpot}
\end{equation}%
with the values of the parameters $\alpha _{i}=\left\{ 0.1,0.55,0.35\right\}
$, $\chi _{i}=\left\{ 6.0,1.2,0.3\right\} $, and with $R_{j}$ being the
screening radius of the $j$-th atom in the elementary cell. The calculations
are carried out for the sinusoidal transversal acoustic wave of the $S$-type
(for the corresponding parameters see, for instance, Ref. \cite{Shas}). For
this wave the vector of the amplitude of the displacement is directed along
the $X$-direction of the quartz single crystal, $\mathbf{u}_{0}\mathbf{=}%
\left( u_{0},0,0\right) $, and the velocity is equal $4.687\cdot 10^{5}$
cm/sec. The vector determining the direction of the hypersound propagation
lies in the $YZ$-plane and forms the angle 0.295 rad with the $Z$-axis. As
the $z$-axis we choose the axis $Z$ of the quartz crystal. The corresponding
function $F\left( x\right) $ is determined by formula (\ref{FmBessel}). The
numerical calculations show that, in dependence of the values for the
parameters $E_{1}$, $\theta $, $\alpha $, $u_{0}$, $\lambda _{s}$, the
external excitation can either enhance or reduce the cross-section of the
bremsstrahlung process.

\begin{figure}[tbph]
\begin{center}
\begin{tabular}{cc}
\epsfig{figure=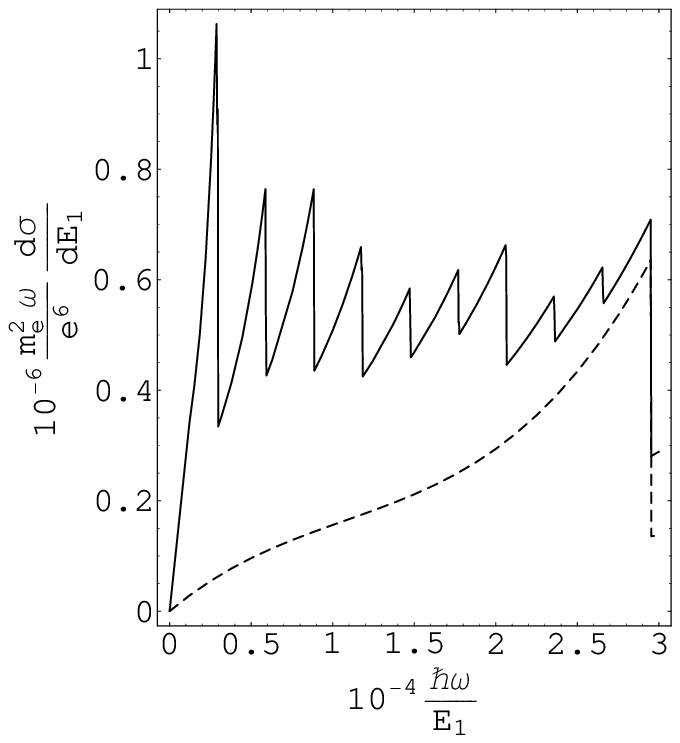,width=6.5cm,height=7cm} & \quad %
\epsfig{figure=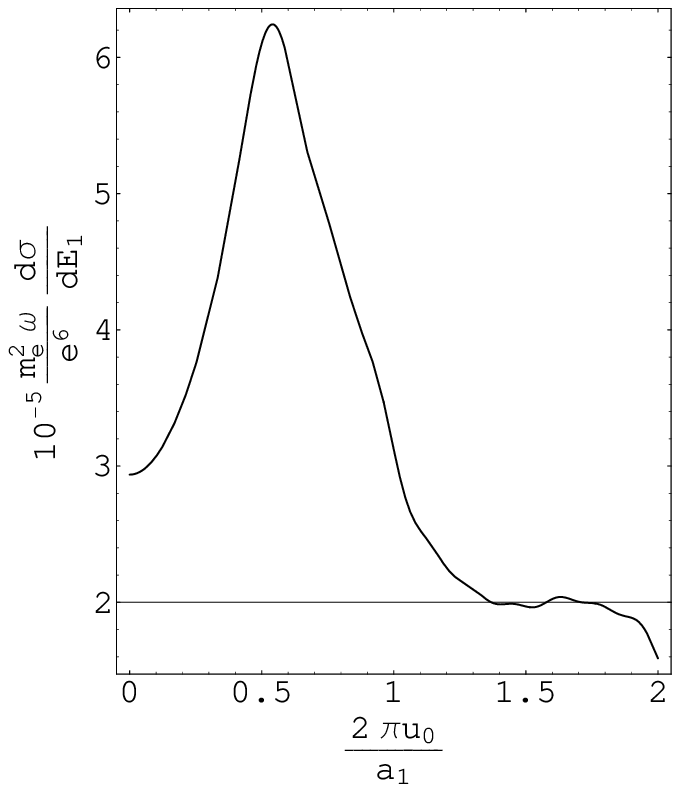,width=6.5cm,height=7cm}%
\end{tabular}%
\end{center}
\caption{The coherent part of the cross-section evaluated by
formula (\protect\ref{sigsumgyintgxstruc}), as a function of $
\protect\omega /E_{1}$ (left panel) for the electron energy
$E_{1}=70$ MeV and $\protect\theta =0.00024$ in the cases
$u_{0}=0$ (dashed curve), $2 \protect\pi u_{0}/a_{2}=0.55$ (full
curve), and as a function of $2\protect \pi u_{0}/a_{2}$ (right
panel) for the photon energy $\protect\omega /E_{1}=0.0002$. The
deformation field is generated by the acoustic wave of the
$S$-type with the frequency $\protect\nu _{s}=5$ GHz.}
\label{fig1}
\end{figure}

\begin{figure}[tbph]
\begin{center}
\begin{tabular}{cc}
\epsfig{figure=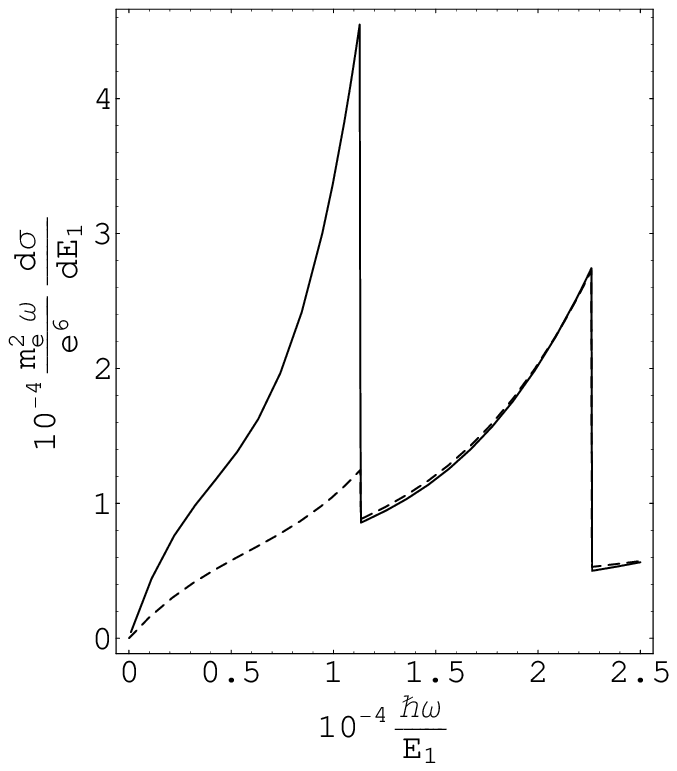,width=6.5cm,height=6.5cm} & \quad %
\epsfig{figure=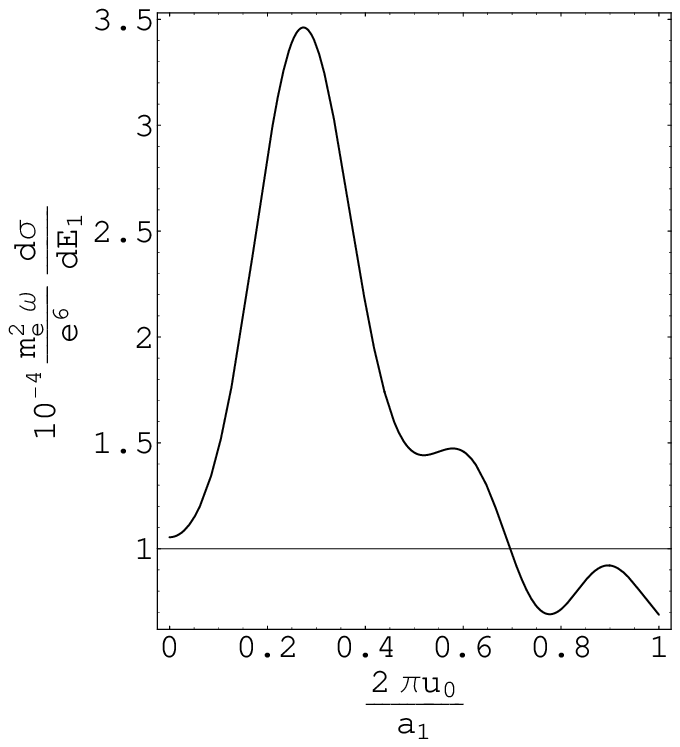,width=6.5cm,height=7cm}%
\end{tabular}%
\end{center}
\caption{The cross-section $10^{-4}\left( m_{e}^{2}\protect\omega %
/e^{6}\right) d\protect\sigma _{c}/d\protect\omega $, evaluated by formula (%
\protect\ref{sigsummn1}), as a function of $\protect\omega /E_{1}$
(left panel) for the electron energy $E_{1}=70$ MeV and
$\protect\psi =0.00092$ in cases $u_{0}=0$ (dashed curve),
$2\protect\pi u_{0}/a_{1}=0.25 $ (full curve) and as a function of
$2\protect\pi u_{0}/a_{1}$ (right panel) for the photon energy
$\protect\omega /E_{1}=0.0001$. } \label{fig2}
\end{figure}

As an illustration of the enhancement for the coherent bremsstrahlung
cross-section, in the left panel of figure \ref{fig1} we have plotted the
quantity $10^{-6}\left( m_{e}^{2}\omega /e^{6}\right) d\sigma _{c}/d\omega $%
, evaluated by formula (\ref{sigsumgyintgxstruc}), as a function of the
ratio $\omega /E_{1}$, for $u_{0}=0$ (dashed curve, deformation is absent)
and for $2\pi u_{0}/a_{2}=0.55$ (full curve) in the case $\theta =0.00024$
and for the electron of energy $E_{1}=70$ MeV moving in $\mathrm{SiO}_{2}$
single crystal. The corresponding deformation field is generated by the
transversal acoustic wave of the $S$-type with the frequency $\nu _{s}=5$
GHz. In the right panel of figure \ref{fig1} the cross-section $%
10^{-6}\left( m_{e}^{2}\omega /e^{6}\right) d\sigma _{c}/d\omega $,
evaluated by formula (\ref{sigsumgyintgxstruc}), is presented as a function
of the relative amplitude of the deformation, parameter $2\pi u_{0}/a_{2}$,
for the energy of photon corresponding to $\omega /E_{1}=0.0002$. The values
of the other parameters are the same as those for the left panel. In the
left panel of figure \ref{fig2} we have presented the cross-section $%
10^{-4}\left( m_{e}^{2}\omega /e^{6}\right) d\sigma _{c}/d\omega $,
evaluated by formula (\ref{sigsummn1}), as a function of the ratio $\omega
/E_{1}$ for $u_{0}=0$ (dashed curve) and $2\pi u_{0}/a_{1}=0.25$ (full
curve) in the case $\psi =0.00092$. The values of the other parameters are
the same as those for figure \ref{fig1}. In the right panel of figure \ref%
{fig2} we have plotted the cross section $10^{-4}\left( m_{e}^{2}\omega
/e^{6}\right) d\sigma _{c}/d\omega $, evaluated by formula (\ref{sigsummn1}%
), as a function of the parameter $2\pi u_{0}/a_{1}$ for $\omega
/E_{1}=0.0001$ with the same values of the other parameters as those for
figure \ref{fig1}.

As we see from the presented examples, the presence of the
deformation field leads to the appearance of additional peaks in
the spectral distribution of the radiated photons for frequencies
lower than the frequency of the first peak in the situation where
the deformation is absent. As we have already mentioned before,
this is related to that in the presence of the deformation field
the number of possibilities to satisfy the condition
$g_{m\parallel }\geqslant \delta $ in the summation in formula
(\ref{sigcohgener}) increases due to the presence of the
additional term $mk_{s\parallel }$ in the expression for
$g_{m\parallel }$.

\section{Conclusion}

\label{sec4}

In the present paper we have investigated the influence of the deformation
field of an arbitrary profile on the coherent bremsstrahlung of high energy
electrons in oriented single crystals with complex base. The latter can be
induced, for example, by acoustic waves. The corresponding results show that
the deformation field can serve as a possible mechanism for the control of
the angular-frequency characteristics of the radiated photons. The
calculations have been done within the framework of the first Born
approximation in the crystal potential. In the presence of periodic
deformation field, the process of coherent bremsstrahlung can be considered
as a result of the electron motion in the continuous periodic potential
given by formula (\ref{EffPot}). The coherent part of the cross-section,
averaged on thermal fluctuations of atoms, is given by formula (\ref%
{sigcohgener}) where the factor $\left\vert F_{m}\left( \mathbf{g}_{m}%
\mathbf{u}_{0}\right) \right\vert ^{2}$ is determined by the function
describing the displacement of the atoms due to the deformation field, and
the factor $\left\vert S\left( \mathbf{g}_{m},\mathbf{g}\right) \right\vert
^{2}$ is determined by the structure of the crystal cell. In addition to the
summation over the reciprocal lattice vector of the crystal, formula (\ref%
{sigcohgener}) contains a summation over the reciprocal lattice vector $m%
\mathbf{k}_{s}$ of the one-dimensional super-lattice induced by the
deformation field. The influence of the deformation field on the
cross-section can be remarkable under the condition $4\pi ^{2}u_{0}/a\gtrsim
\lambda _{s}/l_{c}$. The role of coherence effects in the bremsstrahlung
cross-section is essential when the electron enters into the crystal at
small angles with respect to the crystallographic axes. In this case the
main contribution into the coherent part of the cross-section comes from the
crystallographic planes, parallel to the chosen axis (axis $z$ in our
consideration). The behavior of the cross-section as a function of the
photon energy essentially depends on the angle $\alpha $\ between the
projection of the electron momentum on the plane $\left( x,y\right) $\ and $y
$-axis. If the electron moves far from the corresponding crystallographic
planes, the summation over the perpendicular components of the reciprocal
lattice vector can be replaced by the integration and the coherent part of
the bremsstrahlung cross-section is given by formula (\ref{sigcohintegrgxgy}%
). When the electron enters into the crystal near crystallographic planes,
two cases have to be distinguished. For the first one $\theta \sim
a_{2}/2\pi l_{c}$, the summation over $g_{x}$ can be replaced by the
integration and one obtains formula (\ref{sigcohintegrgxsumgy}). This
formula is simplified to the form (\ref{sigsumgyintgxstruc}) in the case
when the amplitude of the deformation field is perpendicular to the
crystallographic $x$-axis. In the second case one has $\psi =\alpha \theta
\sim a_{1}/2\pi l_{c}$,\ and the main contribution into the cross-section
comes from the crystallographic planes parallel to the incidence plane. The
corresponding formula for the cross-section takes the form (\ref{sigsummn1}%
). The numerical calculations for the cross-section are carried for the $%
\mathrm{SiO}_{2}$\ single crystal with the Moliere parametrization
of the screening atomic potentials and for the deformation field
generated by the transversal acoustic wave of the $S$-type with
frequencies $5$ GHz, and for the energy of an electron $70$ MeV.
Results of the numerical calculations are presented in figures
\ref{fig1} and \ref{fig2}. The values of the parameters $\theta $,
$\psi $, $u_{0}$ are chosen in the way to have an enhancement of
the cross-section for the bremsstrahlung. As it is seen from the
given figures, the presence of an ultrasonic wave leads to the
appearance of new peaks in the cross-section of a coherent
bremsstrahlung. This is related to that in the presence of the
ultrasonic waves the number of possibilities to satisfy the
condition $g_{m\parallel }\geqslant \delta $ increases. These new
peaks are relatively strong in the range of the ratio $\omega
/E_{1}$ from zero up to the first peak of the cross-section in the
case when the ultrasonic wave is absent.

\end{document}